\documentclass[%
 reprint,
 amsmath,amssymb,
aps
]{revtex4-1}
\usepackage{url}
\usepackage[colorlinks=true, linkcolor=blue,urlcolor=blue,anchorcolor=blue,citecolor=blue,bookmarksnumbered]{hyperref}
\usepackage{graphicx}
\usepackage{dcolumn}
\usepackage{bm}

\begin{document}

\title{Unselective ground-state blockade of Rydberg atoms for implementing quantum gates}
\author{Jin-Lei Wu$^{1}$}\author{Yan Wang$^{1}$}\author{Jin-Xuan Han$^{1}$}\author{Shi-Lei Su$^{2}$}\author{Yan Xia$^{3}$}\author{Yongyuan Jiang$^{1}$}\author{Jie Song$^{1,4,5}$}\email[]{jsong@hit.edu.cn}

\affiliation{$^{1}$School of Physics, Harbin Institute of Technology, Harbin 150001, China}
\affiliation{$^{2}$School of Physics and Microelectronics, Zhengzhou University, Zhengzhou 450001, China}
\affiliation{$^{3}$Department of Physics, Fuzhou University, Fuzhou 350002, China}

\begin{abstract}
A dynamics regime of Rydberg atoms, unselective ground-state blockade~(UGSB), is proposed in the context of Rydberg antiblockade~(RAB), where the evolution of two atoms is suppressed when they populate in an identical ground state. UGSB is used to implement a SWAP gate in one step without individual addressing of atoms. Aiming at circumventing common issues in RAB-based gates including atomic decay, Doppler dephasing, and fluctuations in the interatomic coupling strength, we modify the RAB condition to achieve a dynamical SWAP gate whose robustness is much greater than that of the nonadiabatic holonomic one in the conventional RAB regime. In addition, on the basis of the proposed SWAP gates, we further investigate the implementation of a three-atom Fredkin gate by combining Rydberg blockade and RAB. The present work may facilitate to implement the RAB-based gates of strongly coupled atoms in experiment.
\end{abstract}
\maketitle

\section{Introduction}
Since the pioneering work of Jaksch et al.~\cite{Jaksch2000} was reported for implementing Rydberg-mediated quantum logic using neutral atoms, cryogenic Rydberg atoms have been considered as a promising candidate platform for quantum computation, due to interatomic long-range, powerful interactions and relatively long coherence time of Rydberg states~\cite{Gallagher1994,Saffman2010}. Owing to the powerful Rydberg–Rydberg interaction~(RRI), the Rydberg excitation of an atom will cause large energy shifts of its adjacent atoms and thus prohibit their resonant Rydberg excitation, that is, Rydberg blockade~\cite{Jaksch2000,Lukin2001,Gallagher1994,Saffman2010}. As one of the most representative phenomena observed in neutral atomic systems~\cite{Urban2009,Gaetan2009}, Rydberg blockade has facilitated various quantum gate schemes proposed for quantum computation~\cite{Moller2008,Wu2010,Wu2017,Zhao2017,Kang2018,Petrosyan2017,Beterov2018,Shi2018,Shen:19,Liao:19,Liu2020,Khazali2020,Saffman2020,Mitra2020,Guo2020,XFShi2020TSD,XFShi2021FOP}. In contrast to the Rydberg blockade, Rydberg antiblokade~(RAB)~\cite{Ates2007,Pohl2009,Qian2009,Amthor2010,Li2013,SLSu2020EPL} provides another way of quantum gates~\cite{Su2016,Su2017,Su2017PRA,Su2018,Wu2020PLA,Xing2020,FQGuo2020}. Making use of the RAB is more direct to perform a quantum gate, since on the one hand the laser fields are imposed on two atoms identically without individual atomic addressing, and on the other hand the conditional operations on two atoms do not needs to keep one atom in a lossy  Rydebrg state for a while of a $2\pi$ pulse. Besides, two- or multi-atom Rydberg excitation enables RAB to become an effective method for generating steady entanglement using dissipation of Rydberg atoms~\cite{Rao2013,Carr2013,XQShao2014,Su2015,Song2017,XQShao2017,Zhu2020}. Based on Rydberg blockade, two-atom entangling gates are being improved with fidelities from slightly larger than $0.5$ initially~\cite{Wilk2010,Isenhower2010,Zhang2010} to higher than $0.95$ recently~\cite{Levine2018,Levine2019,Madjarov2020}. As for RAB-based entangling gates, besides a two-atom entanglement in the weak coupling regime with a fidelity $0.59$ reported lately~\cite{Jo2020}, an experimental demonstration for strongly coupled atoms are still in anticipation.

In addition to Rydberg blockade and antiblockade, there are other dynamics regimes that are useful for exploring properties of interacting Rydberg atoms or implementing some certain quantum tasks, such as Rydberg dressing~\cite{Balewski2014,Jau2015}, unconventional Rydberg pumping~\cite{DXLi2018}, ground-state blockade~(GSB) of Rydberg atoms~\cite{Shao2017GSB}, and so on. Among these dynamics regimes, GSB of Rydberg atoms is proposed between two $N$-type Rydberg atoms by combining the RAB effect and the Raman transition by using three laser fields~\cite{Shao2017GSB}. By using RAB, two atoms can be excited simultaneously from a collective ground state $|ee\rangle$ to the Rydberg pair state $|rr\rangle$. This doubly-excited Rydberg pumping is set to be very strong so as to induce the quantum Zeno effect, where this relatively strong RAB interaction acts as a measuring device to observe frequently the evolution of $|ee\rangle$ and thus to suppress the evolution of $|ee\rangle$. GSB can block selectively the system evolution of one two-atom ground states, which has been applied to generate entangled states~\cite{Shao2017GSB,YJZhao2017,YHChen2018,Li:19} and has been generalized to an alternative format with a Rydberg dipole-dipole interaction~\cite{Shao2020}.

In this work, we propose a dynamics regime of Rydberg atoms, termed ``unselective ground-state blockade''~(UGSB). In this regime, the evolution will be blocked as long as the two atoms populate in an identical ground state, so through an effective Raman-like process $|01\rangle\leftrightarrow|rr\rangle\leftrightarrow|10\rangle$, a straightforward nonadiabatic holonomic SWAP gate can be implemented. Different from the existing SWAP gate schemes of Rydberg atoms that involve two Rydberg states in an individual atom or needs multi-step operations~\cite{Wu_2012,Shi2014,Glaetzle2015}, the present SWAP gate scheme involves solely one Rydberg state in an individual atom, and is conducted in one step without individual atomic addressing. There are some common issues in RAB-based gates~\cite{Su2016,Su2017,Su2017PRA,Su2018,Wu2020PLA,Xing2020}, which also appear in the present SWAP gate scheme, including significant effect of atomic decay, fragile resilience against Doppler dephasing, and extreme sensitivity to deviation of interatomic interaction strength. To solve these issues, we modify the condition of constructing RAB to implement a dynamical SWAP gate, based on which the gate robustness is significantly strengthened. The present work shows a valid method for circumventing the common issues in the RAB-based gates, and hence may greatly promote the experimental demonstrations of the RAB-based gates of strongly coupled atoms.

\begin{figure}[t]\centering
	\includegraphics[width=\linewidth]{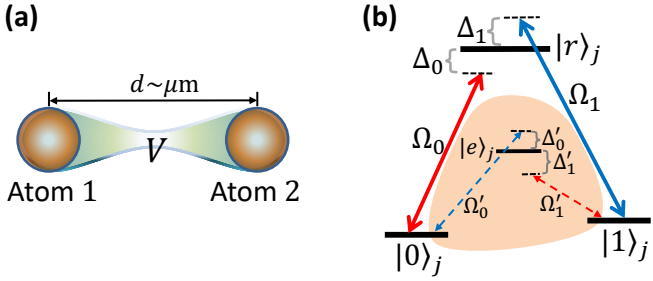}\\ 
	\caption{Schematic for constructing UGSB of two Rydberg atoms and for implementing a SWAP gate. (a)~Two identical atoms 1 and 2 are coupled to each other through RRI with strength $V=C_6/d^6$, $C_6$ being the van der Waals interaction coefficient and $d$ the interatomic distance of several microns. (b)~Two ground states in each atom are encoded as qubit states $|0\rangle$ and  $|1\rangle$ that are pumped into a Rydberg state $|r\rangle$ through two off-resonant laser fields, respectively, with Rabi frequencies $\Omega_0$ and $\Omega_1$, corresponding to respective red detuning $\Delta_0$ and blue detuning $\Delta_1$. Shaded area: auxiliary atom-field interactions induce additional Stark shifts of $|0\rangle_j$ and $|1\rangle_j$ to offset ones caused by detuned Rydberg pumping fields. The ground state $|0\rangle$~($|1\rangle$) is coupled to an excited state $|e\rangle$ of a lower principal quantum number with Rabi frequency $\Omega'_0$~($\Omega'_1$) and blue detuning $\Delta'_0$~(red detuning $\Delta'_1$). $|e\rangle$ is virtually excited with $\Delta'_k\gg\Omega'_k$~($k=0,1$). The auxiliary fields and Rydberg pumping fields are synchronously imposed.}\label{f1}
\end{figure}
Furthermore, based on the proposed SWAP gates, the implementations of three-atom controlled-SWAP~(Fredkin) gates are explored by combining Rydberg blockade and antiblockade. Now that the effective system for the SWAP gate on the two (target) atoms can be regarded as a three-level Rydberg atom \textit{in form} with a Rydberg state $|rr\rangle$ and two ground states $|01\rangle$ and $|10\rangle$, excitation of an introduced (control) Rydberg atom will suppress excitation of the two target atoms because of Rydberg blockade. Therefore, after exciting the control atom from a ground state~($|0\rangle$) to the Rydberg state, the SWAP operation on the target atoms can not be implemented. However, when the control atom populates in $|1\rangle$ that can not be excited, the SWAP operation works well on the target atoms. The process above indicates the implementation of a Fredkin gate where the SWAP operation on the two target atoms depends on the state of the control atom. Compared with the recent work~\cite{Wu2021PR} that focuses on the one-step achievements of Rydberg SWAP and also three-qubit controlled-SWAP gates, for the SWAP gate the present work does not need modulations in the amplitudes of drive fields and provides instructional methods for eliminating unwanted shifts of energy levels. As for the Fredkin gate, the present three-step scheme combines Rydberg blockade and antiblockade, which does not require a fixed value of the sum RRI strength between the control atom and the target ones. Therefore, the present Fredkin gate is more robust to errors in RRI strengths than that in Ref.~\cite{Wu2021PR} which requires a fixed relation between the control-target RRI strength and the modulation frequency of the field on the control atom.

This paper is organized as follows: in Sec.~\ref{sec2}, we illustrate the construction of UGSB using two Rydberg atoms. In Sec.~\ref{sec3}, we apply UGSB to implement SWAP gates, including the nonadiabatic holonomic and the dynamical versions. In Sec.~\ref{sec4}, the robustness of two kinds of SWAP gates are studied and compared. In Sec.~\ref{sec5}, based on two kinds of SWAP gates, we propose to implement Fredkin gates. Finally, a conclusion is given in Sec.~\ref{sec6}.

\section{Construction of unselective ground-state blockade of Rydberg atoms}\label{sec2}
\subsection{Description of two interacting Rydberg atoms}
\begin{figure}[t]\centering
	\includegraphics[width=\linewidth]{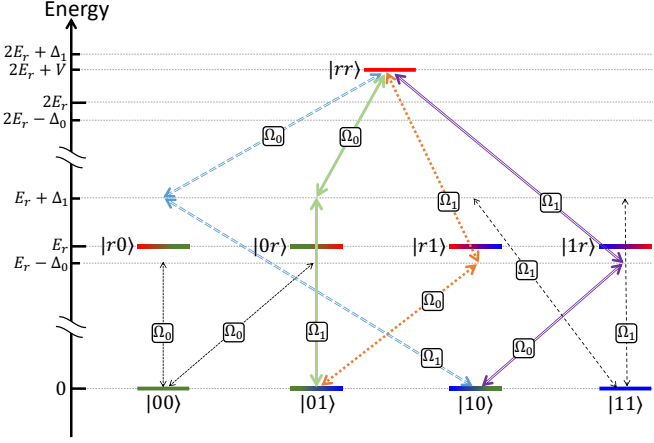}\\ 
	\caption{Level transitions of two-atom collective states with considering comprehensively the laser-atom interaction and the atom-atom interaction. Energy of atomic ground states are set as zero. Energy of the Rydberg state is assumed as $E_r$. $\Delta_0,\Delta_1,V>0$ is assumed.}\label{f2}
\end{figure}

As shown in Fig.~\ref{f1}(a), two Rydberg atoms with laser-driven level transitions shown in Fig.~\ref{f1}(b)~(the shaded area is not considered for the time being) are coupled to each other through RRI. The Hamiltonian of this two-atom system in the Schrödinger picture with the rotating-wave approximation reads~($\hbar=1$)
\begin{eqnarray}\label{e1}
	\hat H_S&=& \hat{H}_0+\hat{H}_i +V|rr\rangle\langle rr|,\nonumber\\
	\hat{H}_0&=&\sum_{j=1}^2\omega_0|0\rangle_j\langle 0|+\omega_1|1\rangle_j\langle 1|+\omega_r|r\rangle_j\langle r|,\nonumber\\
	\hat{H}_i&=&\sum_{j=1}^2\frac{\Omega_0}2|0\rangle_j\langle r|e^{{\rm i}\omega_{l0}t}+\frac{\Omega_1}2|1\rangle_j\langle r|e^{{\rm i}\omega_{l1}t}+{\rm H.c.},
\end{eqnarray}
where $\omega_m$~($m=0,1,r$) describes the frequency of the atomic
level $|m\rangle$ and $\omega_{lk}$~($k=0,1$) represents the center frequency
of the laser field driving the atomic transition $|k\rangle\leftrightarrow|r\rangle$. $V|rr\rangle\langle rr|$ with $|rr\rangle\equiv|r\rangle_1\otimes|r\rangle_2$ denotes the energy shift of the Rydberg pair state caused by the RRI between the atoms.
The corresponding Hamiltonian in the interaction picture can be obtained
\begin{equation}\label{e2}
	\hat H_I= \hat{\mathcal{H}}_1+\hat{\mathcal{H}}_2 +V|rr\rangle\langle rr|,
\end{equation}
where  $\hat{\mathcal{H}}_j$~($j=1,2$) is individual Hamiltonian of the $j$-th atom driven by Rydberg pumping fields
\begin{equation}\label{e3}
	\hat{\mathcal{H}}_j=\sum_{k=0}^1\frac{\Omega_k}{2}|r\rangle_j\langle k|e^{(-1)^k{\rm i}\Delta_kt}+{\rm H.c.}
\end{equation}
in which we have defined detunings as $\Delta_k\equiv(-1)^k(\omega_r-\omega_k-\omega_{lk})$ to make $\Delta_k$ positive for convenience of discussion in the following.
In order to see joint results of the laser-atom interaction and the atom-atom interaction, it is convenient to transform the system into the frame defined by a rotation operator $U_0=\exp({\rm i}tV|rr\rangle\langle rr|)$, and then the two-atom Hamiltonian is written with the two-atom basis \{$|pq\rangle$\}~($p,q=0,1,r$) as
\begin{eqnarray}\label{e4}
	&&\hat{\mathcal{H}}_I=\nonumber\\
	&&\frac{\Omega_0}2e^{-{\rm i}\Delta_0t}(|00\rangle\langle r0|+|01\rangle\langle r1|+|00\rangle\langle 0r|+|10\rangle\langle 1r|)\nonumber\\
	&&+\frac{\Omega_1}2e^{{\rm i}\Delta_1t}(|10\rangle\langle r0|+|11\rangle\langle r1|+|01\rangle\langle 0r|+|11\rangle\langle 1r|)\nonumber\\
	&&+\frac{\Omega_0}2e^{-{\rm i}(\Delta_0+V)t}(|0r\rangle\langle rr|+|r0\rangle\langle rr|)\nonumber\\
	&&+\frac{\Omega_1}2e^{{\rm i}(\Delta_1-V)t}(|1r\rangle\langle rr|+|r1\rangle\langle rr|)+{\rm H.c.}
\end{eqnarray}
This Hamiltonian can be analyzed by visualizing it with level transitions, as shown in Fig.~\ref{f2}.

\subsection{Effective dynamics for unselective ground-state blockade}
When considering the condition $\Delta_{0,1},|\Delta_1-V|\gg|\Omega_{0,1}|/2$, all one-photon transitions between the single-excitation states and the ground states or the double-excitation state $|rr\rangle$ are largely detuned. Therefore, all single-excitation states~(i.e., $|0r\rangle$, $|1r\rangle$, $|r0\rangle$, and $|r1\rangle$) can be adiabatically eliminated. However, it is possible to make the two-photon transitions between $|rr\rangle$ and some certain ground states fully or nearly resonant by assigning appropriate relation among $\Delta_0$, $\Delta_1$, and $V$. We intuitively assume a relation $\Delta_1-\Delta_0=V$ that ensures the resonant two-photon transitions between $|rr\rangle$ and $|01\rangle$~($|10\rangle$) mediated by $|r1\rangle$ and $|0r\rangle$~($|r0\rangle$ and $|1r\rangle$), but excludes transitions between $|rr\rangle$ and $|00\rangle$~($|11\rangle$). According to the second-order perturbation theory, an effective form of Hamiltonian Eq.~(\ref{e4}) can be obtained~(See Appendix~\ref{Appendix} for details)
\begin{eqnarray}\label{e5}
	\hat{\mathcal{H}}'_I &=&\sum_{m,n=0,1}\Delta_{mn}|mn\rangle\langle mn|+\Delta_{rr}|rr\rangle\langle rr|\nonumber\\
	&&+\Big[\frac{\Omega_{\rm eff}}2(|01\rangle\langle rr|+|10\rangle\langle rr|)+{\rm H.c.}\Big],
\end{eqnarray}
with
\begin{eqnarray}\label{e6}
	&&\Delta_{00}=-\frac{\Omega_0^2}{2\Delta_0},\quad\Delta_{01}=\Delta_{10}=\frac{\Omega_1^2}{4\Delta_1}-\frac{\Omega_0^2}{4\Delta_0},\nonumber\\
	&&\Delta_{11}=\frac{\Omega_1^2}{2\Delta_1},\quad\Delta_{rr}=\frac{\Omega_0^2}{2\Delta_1}-\frac{\Omega_1^2}{2\Delta_0},\nonumber\\
	&&\Omega_{\rm eff}=\frac{\Omega_0\Omega_1}{2\Delta_1}-\frac{\Omega_0\Omega_1}{2\Delta_0}.
\end{eqnarray}
Terms in the first line of Eq.~(\ref{e5}) denote the Stark shifts of four ground product states and the doubly-excited Rydberg pair state $|rr\rangle$, while the last term is an effective coupling $|01\rangle\leftrightarrow|rr\rangle \leftrightarrow|10\rangle$.

Here we introduce opposite-sign energy shifts caused by auxiliary detuned transitions to offset the unwanted ground-state Stark shifts in Eq.~(\ref{e5}). The auxiliary detuned transitions are shown in the shaded area of Fig.~\ref{f1}(b). The auxiliary atom-field interaction for the $j$-th~($j=1,2$) atom is described by Hamiltonian  $\hat H'_j=\sum_{k=0}^1{\Omega'_k}/{2}|k\rangle_j\langle e|e^{(-1)^k{\rm i}\Delta'_kt}+{\rm H.c.}$ with $\Delta'_k\gg\Omega'_k$~($k=0,1$), only leading to energy shifts of ground states that exactly offset counterparts induced by the Rydberg pumping fields. These two auxiliary fields are supposed to be synchronous 
with the Rydberg pumping fields so as to avoid superfluous phase rotations of the ground states, where $\Omega'_k=\Omega_j$ and $\Delta'_k=\Delta_k$~($k=0,1$) are assumed hereinafter for simplicity. As for the Stark shift of $|rr\rangle$, it can be absorbed by changing the RAB condition from $V-(\Delta_1-\Delta_0)=0$ into $V-(\Delta_1-\Delta_0)=V_0$ with $V_0$ being a relatively small quantity compared with $\Delta_{0,1}$ and $V$. Hence a final effective Hamiltonian is attained~(See Appendix~\ref{A2} for dealing with the Stark shifts in details)
\begin{equation}\label{e7}
	\hat H_{\rm eff}=\Big[\frac{\Omega_{\rm eff}}2(|01\rangle\langle rr|+|10\rangle\langle rr|)+{\rm H.c.}\Big]+\delta|rr\rangle\langle rr|,
\end{equation}
for which $\delta\equiv V_0+\Delta_{rr}$ can be assigned as needed.
The effective quantum system described by Eq.~(\ref{e7}) indicates the UGSB regime, where the evolution from even-parity states $|00\rangle$ and $|11\rangle$ is blocked, while the doubly-excited Rydberg pair state $|rr\rangle$ mediates the transition between two odd-parity computational states $|01\rangle$ and $|10\rangle$. The underlying physics of the UGSB regime originates from the interference of the opposite-sign detunings of the two transitions $|0\rangle\leftrightarrow|r\rangle$ and $|1\rangle\leftrightarrow|r\rangle$ of the two atoms. 
When two atoms both populate in $|0\rangle$~($|1\rangle$), the detuning of the transition from the two-atom state $|00\rangle$~($|11\rangle$) to $|rr\rangle$ will be doubled in addition to the RRI, which thus blocks the transition of $|00\rangle$ and $|11\rangle$. However, the detuning of the transition from the two-atom state $|01\rangle$~($|10\rangle$) to $|rr\rangle$ will be complementary to neutralize the RRI, which can exactly make the transition $|01\rangle~(|10\rangle)\leftrightarrow|rr\rangle$ resonant.

\begin{figure*}\centering
	\includegraphics[width=0.9\linewidth]{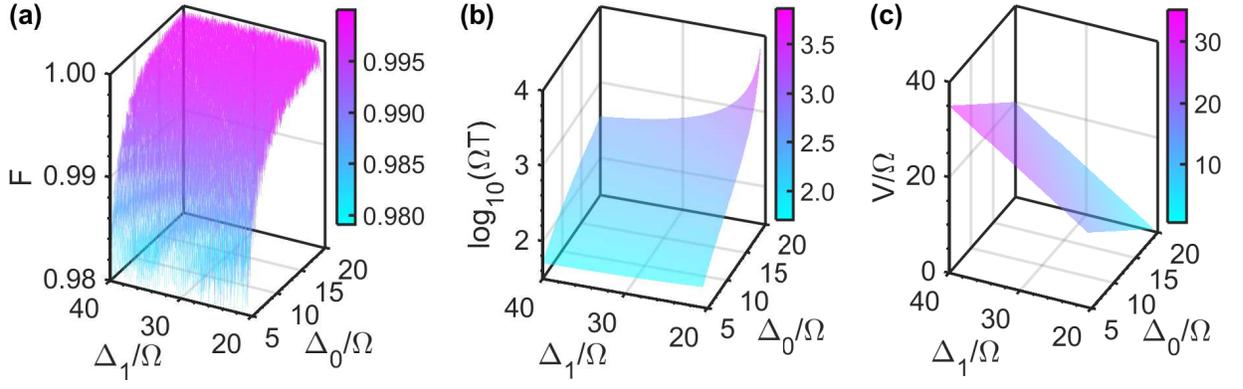}\\ 
	\caption{(a)~Fidelity at the time $t=T$ with different $\Delta_0/\Omega\in[5,~20]$ and $\Delta_1/\Omega\in[20.1,~40]$ of performing the SWAP gate on the initial state $(|0\rangle_1+|1\rangle_1)/\sqrt2\otimes(|0\rangle_2+|1\rangle_2)/\sqrt2$. (b)~Gate time parameter $\log_{10}(\Omega T)$ with different $\Delta_0/\Omega$ and $\Delta_1/\Omega$. (c)~RRI strength parameter $V/\Omega$ with different $\Delta_0/\Omega$ and $\Delta_1/\Omega$.}\label{f3}
\end{figure*}
\section{Implementation of SWAP gates}\label{sec3}
\subsection{Nonadiabatic holonomic SWAP gate}
Nonadiabatic holonomic quantum computation (NHQC) was proposed for implementing fast and robust gate operations in three-level systems~\cite{Sj_qvist_2012,GFXu2012,GRFeng2013}, which is promising for fault-tolerant quantum computation due to the built-in resilience of geometric phases to certain local noises. For a clear illustration, we recall the basic concept of NHQC~\cite{Sj_qvist_2012,GFXu2012,Kang2018,PZZhao2020}. An $N$-dimensional quantum system with Hamiltonian $\hat{H}$ and an $L$-dimensional time-dependent subspace $\mathcal{S}(t)$ spanned by a set of orthonormal basis $\{|\phi_l(t)\rangle\}^L_1$ are considered,	where $|\phi_l(t)\rangle$ is the solution of the Schr\"odinger equation ${\rm i}\partial_t|\phi_l(t)\rangle=\hat H(t)|\phi_l(t)\rangle$. A nonadiabatic holonomic gate needs to satisfy two conditions:
\begin{eqnarray*}
	&&({\rm i})~\sum_{l=1}^L|\phi_l(T)\rangle\langle\phi_l(T)|=\sum_{l=1}^L|\phi_l(0)\rangle\langle\phi_l(0)|,\nonumber\\
	&&({\rm ii})~\langle\phi_l(t)|\hat H|\phi_{l'}(t)\rangle=0,\quad(l,l'=1,2,\cdots,L),
\end{eqnarray*}
where the former is called the cyclic evolution condition while the latter the parallel transport condition.	

Now that even-parity computational states $|00\rangle$ and $|11\rangle$ are not involved in the evolution of two atoms, i.e., $|00\rangle\mapsto|00\rangle$ and $|11\rangle\mapsto|11\rangle$, a SWAP gate can be implemented by exchanging the populations between $|01\rangle$ and $|10\rangle$ through a resonant Raman-like process $|01\rangle\leftrightarrow|rr\rangle\leftrightarrow|10\rangle$ with the resonance condition $\delta =0$.
To this end, Hamiltonian Eq.~(\ref{e7}) can be written as $\hat H'_{\rm eff}=\Omega_{\rm eff}|B\rangle\langle rr|/{\sqrt2}+{\rm H.c.}$ with defining $|B\rangle\equiv(|01\rangle+|10\rangle)/{\sqrt2}$. There also exists an orthogonal state $|D\rangle\equiv(|01\rangle-|10\rangle)/{\sqrt2}$ that is uncoupled to the effective system. Then undergoing a cyclic evolution $|B\rangle\mapsto-|B\rangle$~($|D\rangle\mapsto|D\rangle$) with the operation time $T=\sqrt2\pi/|\Omega_{\rm eff}|$, a nonadiabatic holonomic quantum operation can be realized, which is corresponding to an evolution operator $U(T)=|00\rangle\langle 00|-|B\rangle\langle B|+|D\rangle\langle D|+|11\rangle\langle 11|$. In the computational space, this evolution operator is exactly a SWAP gate $U_{\rm  SWAP}=|00\rangle\langle00|-|01\rangle\langle10|-|10\rangle\langle01|+|11\rangle\langle11|$ that is equivalent to the standard form up to local phase operations, and besides the evolution process satisfies the conditions (i) and (ii) of NHQC.

Here we set $\Omega_0=\Omega_1=\Omega$. In order to ensure the second-order perturbation theory to work well for implementing UGSB and the SWAP gate, the ratios $|\Delta_{0}/\Omega|$ and $|\Delta_{1}/\Omega|$ should be as large as possible. However, the gate time $T$ should be as short as possible to make the system sustain less decoherence. In addition, the RRI strength $V$ should be experimentally feasible. In order to look for suitable  $|\Delta_{0}/\Omega|$ and $|\Delta_{1}/\Omega|$, in Fig.~\ref{f3}(a) we numerically calculate the fidelity at the time $t=T$ with different $|\Delta_{0}/\Omega|$ and $|\Delta_{1}/\Omega|$ of a target state $|\Psi_t\rangle=U_{\rm  SWAP}|\Psi_0\rangle$ with $|\Psi_0\rangle$ being the initial state of two atoms. The fidelity is defined by $F(t)=|\langle \Psi_t|\psi(t)\rangle|^2$ with $|\psi(t)\rangle$ being the solution of Schr\"odinger equation $i\partial |\psi(t)\rangle/\partial t=\hat H_I|\psi(t)\rangle$. Besides, we also calculate values of the gate time $\Omega T$ and the RRI strength $V/\Omega$ with different $|\Delta_{0}/\Omega|$ and $|\Delta_{1}/\Omega|$ in Figs.~\ref{f3}(b) and \ref{f3}(c), respectively. From Fig.~\ref{f3} we learn that the fidelity is over 0.99 with $\Delta_0/\Omega>6$ and $\Delta_1/\Omega>21$. The gate time increases with increasing $\Delta_0/\Omega$ or decreasing $\Delta_1/\Omega$, and the effect of $\Delta_0/\Omega$ on the gate time is more significant than that of $\Delta_1/\Omega$. From Figs.~\ref{f3}(a) and \ref{f3}(c) together we know that a weak or intermediate-scale RRI strength, i.e., $V<\Omega$, can also ensure a high fidelity $F(T)>0.99$, but it consumes a very long gate time. In the following discussion, we choose $\Delta_0=10\Omega$ and $\Delta_1=30\Omega$ that give $F(T)=0.9978$, $\Omega T/2\pi=21.21$, and $V=\Delta_1-\Delta_0+V_0$ with $V_0=\Omega/30$.

\begin{figure*}\centering
	\includegraphics[width=0.7\linewidth]{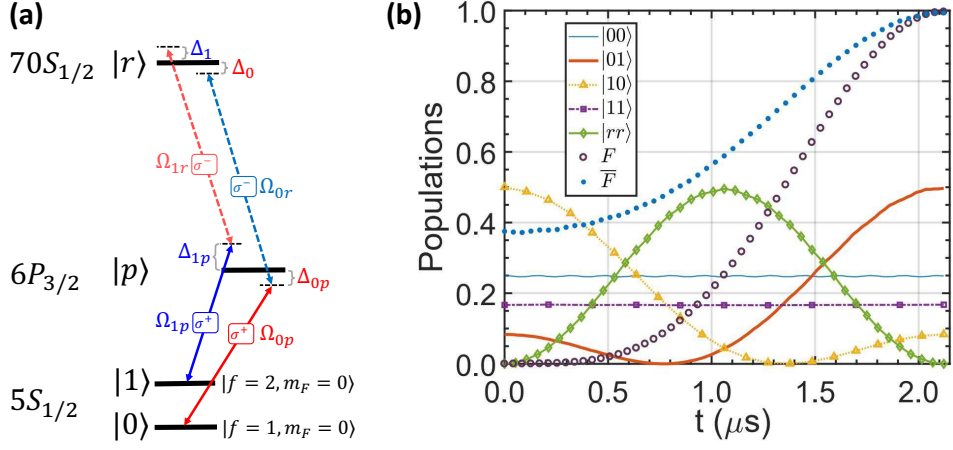}\\ 
	\caption{(a)~Realization of the UGSB transitions through two-photon processes in two $^{87}Rb$ atoms. Qubit states are encoded on two hyperfine ground states $|0\rangle=|5S_{1/2}, F=1, m_{F}=0\rangle$ and $|1\rangle=|5S_{1/2}, F=2, m_{F}=0\rangle$. (b)~Initial-state-specified fidelity of the SWAP gate, populations of the four computational states and the doubly-excited Rydberg pair state, and average fidelity of the SWAP gate with $N=21$, i.e., containing $21^6$ initial states. $\Delta_0/2\pi=100$~MHz, $\Delta_1/2\pi=300$~MHz, and $\Omega_0/2\pi=\Omega_1/2\pi=10$~MHz.}\label{f4}
\end{figure*}
\subsection{Simulation with $^{87}Rb$ atoms}
For realizing the transitions of UGSB in experiment, the Rydberg excitation from the ground states to the Rydberg state can be achieved through two-photon processes in $^{87}Rb$ atoms~\cite{Levine2018,Bernien2017,Levine2019,Omran2019}. Here we encode the qubit states on two hyperfine ground states $|0\rangle=|5S_{1/2}, F=1, m_{F}=0\rangle$ and $|1\rangle=|5S_{1/2}, F=2, m_{F}=0\rangle$ of $^{87}Rb$ atoms. Similar to the recent experiment~\cite{Levine2019},
these two ground states can be excited to a Rydberg state $|r\rangle=|70S_{1/2},m_J=-1/2\rangle$ with $C_6/2\pi=858.4~{\rm GHz}\cdot\mu{\rm m}^6$ by two-photon processes with assistance of an intermediate state $|6P_{3/2}\rangle$, as shown in Fig.~\ref{f4}(a). The off-resonant Rydberg pumping from $|0\rangle$~($|1\rangle$) to $|r\rangle$ with red detuning $\Delta_0$~(blue detuning $\Delta_1$) is achieved by two off-resonant transitions $|0\rangle\leftrightarrow|p\rangle=|6P_{3/2}\rangle$~($|1\rangle\leftrightarrow|p\rangle$) with Rabi frequency $\Omega_{0p}$~($\Omega_{1p}$) and detuning $\Delta_{0p}$~($\Delta_{1p}$) and $|p\rangle\leftrightarrow|r\rangle=|70S_{1/2}\rangle$ with Rabi frequency $\Omega_{0r}$~($\Omega_{1r}$) and detuning $\Delta_{0p}-\Delta_0$~($\Delta_{1p}-\Delta_1$), driven by $\sigma^+$ and $\sigma^-$ polarized lasers, respectively.

We set parameters \{$\Delta_0/2\pi=100$~MHz, $\Delta_{0p}/2\pi=4050$~MHz, $\Omega_{0r}/2\pi=200$~MHz, $\Omega_{0p}/2\pi=400$~MHz\} and \{$\Delta_1/2\pi=300$~MHz, $\Delta_{1p}/2\pi=25151$~MHz, $\Omega_{1r}/2\pi=500$~MHz, $\Omega_{1p}/2\pi=1000$~MHz\}, yielding effective Rabi frequencies $\Omega_k=\Omega_{kp}\Omega_{kr}/2\bar{\Delta}_k=2\pi\times10$~MHz~($k=0,1$) with $\bar{\Delta}_k=2/[1/(\Delta_{kp}-\Delta_{k})+1/\Delta_{kp}]$~\cite{James2007}. The required RRI strength is $V/2\pi=200.33$~MHz according to $V=\Delta_1-\Delta_0-\Delta_{rr}$. For the Rydberg state $|r\rangle=|70S_{1/2},m_J=-1/2\rangle$ with $C_6/2\pi=858.4~{\rm GHz}\cdot\mu{\rm m}^6$~\cite{Bernien2017}, this RRI strength is experimentally feasible with the separation $d=4.03~\mu{\rm m}$ between the atoms. Using this set of parameters, in Fig.~\ref{f4}(b) we plot the fidelity of the SWAP gate and populations of the four computational states and the doubly-excited Rydberg pair state with the initial state $|\Psi_0\rangle=(|0\rangle_1+\sqrt2|1\rangle_1)/\sqrt3\otimes(\sqrt3|0\rangle_2+|1\rangle_2)/2$. It is identified that during the whole evolution the even-parity states $|00\rangle$ and $|11\rangle$ remain unchanged while the population swapping between the odd-parity states $|01\rangle$ and $|10\rangle$ takes place. Therefore, the fidelity of the target state evolves smoothly from $F(0)=0$ initially to $F(T)=0.9966$ finally.

For identifying the randomicity of a two-atom initial state for the nonadiabatic holonomic SWAP gate, we also plot the evolution of an average fidelity for performing the SWAP gate in Fig.~\ref{f4}(b)~(blue-dotted line).
The average fidelity is defined as
\begin{equation}\label{e8}
	\overline F(t)=\frac1{N^6}\sum_{j1}^N\sum_{j2}^N\dots\sum_{j6}^N|\langle\Psi_t|\psi(t)\rangle|^2,
\end{equation}
in which $|\psi(t)\rangle$ is the two-atom state solved by Sch\"{o}dinger equation with an arbitrary initial state $|\psi(0)\rangle=\sin\beta^{(1)}_{j1}|00\rangle+\cos\beta^{(1)}_{j1}[e^{i\beta^{(4)}_{j4}}\sin\beta^{(2)}_{j2}|01\rangle+\cos\beta^{(2)}_{j2}(e^{i\beta^{(5)}_{j5}}\sin\beta^{(3)}_{j3}|10\rangle
+e^{i\beta^{(6)}_{j6}}\cos\beta^{(3)}_{j3}|11\rangle)]$. \{$\beta^{(k)}_{jk}$\}~($k=1,2,\cdots,6$) are six angle parameters, each of which can be assigned with $N$ values uniformly distributed over $(0,~2\pi]$. $|\Psi_t\rangle=U_{\rm SWAP}|\psi(0)\rangle$ is the target state of performing the SWAP gate on the initial state. This definition means the average of fidelities at each moment for $N^6$ input states, and a large enough number $N$ will represent randomicity of input states. The average fidelity shown in Fig.~\ref{f4}(b) can reach $\overline F(T)=0.9936$, which indicates that a high fidelity SWAP gate can be obtained for an arbitrary input state when not considering decoherence.

\subsection{Dynamical SWAP gate with a modified RAB condition}
Geometric-phase-based nonadiabatic holonomic gates are robust against some certain stochastic noises because geometric phases do not rely on evolution details but solely on the geometric trajectory of evolution paths~\cite{BJLiu2019,PZZhao2020}. However, concerning other decoherence factors, such as atomic decay, Doppler dephasing, and deviation in the RRI strength, the nonadiabatic holonomic SWAP gate above is very susceptible, because the Rydberg pair state $|rr\rangle$ attends the evolution. Participation of $|rr\rangle$ is based on the demanding fixed relation among $\Delta_0$, $\Delta_1$, and $V$, causing the strong dependence of the SWAP gate on the precise control of the RRI strength. Besides, the excitation into $|rr\rangle$ will inevitably result in the atomic decay due to finite lifetime of the Rydberg state and in the Doppler dephasing due to finite atomic temperature. These issues are common and 
intractable in almost all of existing RAB-based gate schemes~\cite{Su2016,Su2017,Su2017PRA,Su2018,Wu2020PLA,Xing2020}, greatly limiting the experimental implementation of high-fidelity RAB-based gates.

\begin{figure}[htb]\centering
	\includegraphics[width=0.88\linewidth]{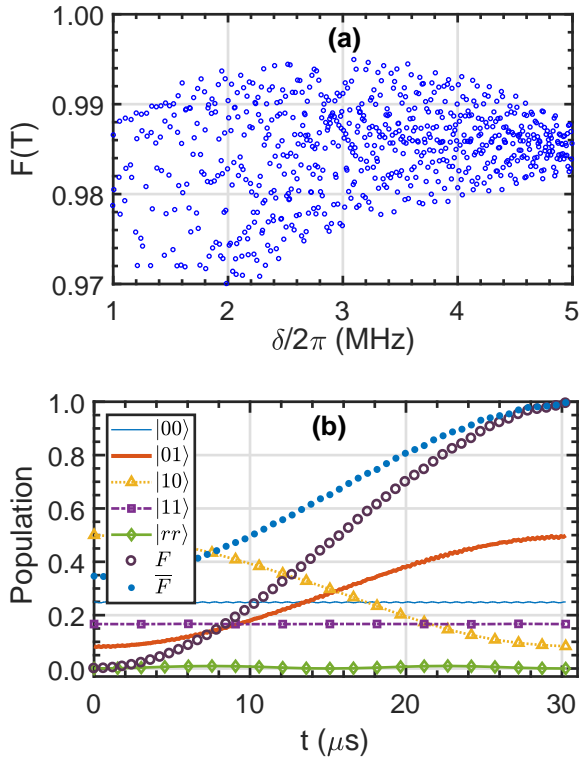}\\ 
	\caption{(a)~Initial-state-specified fidelity of performing the SWAP gate with varying $\delta$. The initial state is $|\Psi_0\rangle=(|0\rangle_1+\sqrt2|1\rangle_1)/\sqrt3\otimes(\sqrt3|0\rangle_2+|1\rangle_2)/2$. (b)~Initial-state-specified fidelity, populations of the four computational states and the doubly-excited Rydberg pair state, and average fidelity of performing the SWAP gate with $N=21$. $\delta/2\pi=3.36$~MHz.}\label{f5}
\end{figure}
In order to solve these issues of RAB-based gates, reducing the participation of
$|rr\rangle$ in the gate procedure can be an effective approach~\cite{Brion_2007,DXLi2018,Wu2020OL,Yin2020,Yin2021}. Here we change the RAB condition for the effective Hamiltonian Eq.~(\ref{e7}) from $\delta=0$ to $|\delta|\gg|\Omega_{\rm eff}|/2$ such that Eq.~(\ref{e7}) is changed into a largely detuned format, which will suppress the attendance of $|rr\rangle$ in the evolution, i.e., avoiding Rydberg excitation. Then an effective interaction between two atoms can be obtained with Hamiltonian
\begin{eqnarray}\label{e9}
	\hat H_{d}&=&-\frac{\Omega_d}2(|01\rangle\langle10|+|10\rangle\langle01|+|01\rangle\langle01|+|10\rangle\langle10|)\nonumber\\
	&=&-{\Omega_{d}}|B\rangle\langle B|,
\end{eqnarray}
with $\Omega_{d}=\Omega_{\rm eff}^2/2\delta$. A dynamical SWAP gate $U_{\rm SWAP}$ can be attained with gate time $T=\pi/|\Omega_{d}|$, which depends on the accumulation of a dynamical $\pi$ phase on $|B\rangle$ with an evolution operator $U_d(T)=\exp({\rm i}T\Omega_{d}|B\rangle\langle B|)$.

In order to choose an appropriate $\delta$ that can ensure a high gate fidelity and give a relatively short gate time, we plot in Fig.~\ref{f5}(a) the initial-state-specified fidelity of performing the SWAP gate with varying $\delta/2\pi$ within $[1,~5]$~MHz. There is a wide range $\delta/2\pi\in[1.5,~4.5]$~MHz containing high fidelities~($F>0.99$). Then in Fig.~\ref{f5}(b) we pick up $\delta/2\pi=3.36$~MHz to plot the initial-state-specified fidelity, populations of the four computational states and the doubly-excited Rydberg pair state, and average fidelity~(blue-dotted line) of performing the dynamical SWAP gate. Apparently, the expected evolution properties of populations in the SWAP gate are identified. Furthermore, the initial-state-specified fidelity and average fidelity indicate that the gate can be achieved with a high fidelity $F(T)=0.9945$ and $\overline F(T)=0.9966$.

\begin{figure}[htb]\centering
	\includegraphics[width=0.88\linewidth]{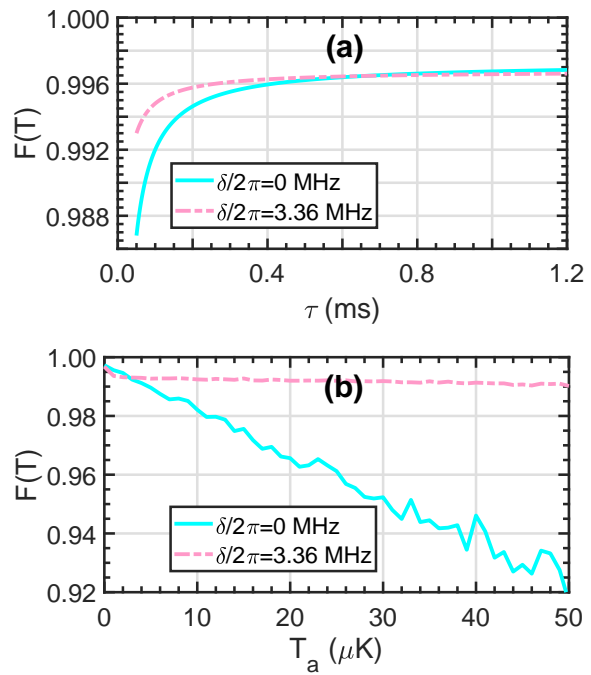}\\ 
	\caption{Fidelities of the nonadiabatic holonomic~($\delta/2\pi=0$) and dynamical~($\delta/2\pi=3.36$~MHz) SWAP gates by considering (a)~atomic decay and (b)~Doppler dephasing. Each point in (b) denotes the average of 201 results that are obtained by picking 201 pairs of random $\delta_{1}$ and $\delta_{2}$ according to a Gaussian probability distribution.}\label{f6}
\end{figure}
\section{Robustness of SWAP gates}\label{sec4}
In this section, we study and compare fidelities of the nonadiabatic holonomic and dynamical SWAP gates by considering atomic decay, Doppler dephasing, and deviations in the RRI strength. The fidelity estimation is based on a specified initial state $|\Psi_0\rangle=(|0\rangle_1+|1\rangle_1)/\sqrt2\otimes|1\rangle_2$, and the intrinsic fidelity without considering any negative factors is $F(T)=0.9973$ and $F(T)=0.9968$, respectively, for the nonadiabatic holonomic and dynamical SWAP gates.

\subsection{Atomic decay}
When atomic decay is taken into account, the evolution of the two-atom system can be dominated by the master equation
\begin{equation}
	{\dot{\rho}}={\rm i}[{\rho},\hat{H}]
	-\frac{1}{2}\sum_{j=1}^2\sum_{k=0}^2\Big(\hat{\mathcal{L}}_{k}^{j\dag}\hat{\mathcal{L}}_{k}^j{\rho}-2\hat{\mathcal{L}}_{k}^j{\rho}\hat{\mathcal{L}}_{k}^{j\dag}+{\rho}\hat{\mathcal{L}}_{k}^{j\dag}\hat{\mathcal{L}}_{k}^j\Big),
\end{equation}
in which ${\rho}$ is the density operator. The atomic decay operator is defined by $\hat{\mathcal{L}}_{k}^j\equiv\sqrt{\gamma_k}|k\rangle_{j}\langle r|$, for which an additional ground state $|2\rangle_j$ is introduced to denote those Zeeman magnetic sublevels out of the qubit states $|0\rangle_j$ and $|1\rangle_j$. In this work, we assume that for $^{87}Rb$ atoms decay rates from a Rydberg state into eight Zeeman ground states are identical for convenience, so $\gamma_0=\gamma_1=1/8\tau$ and $\gamma_2=3/4\tau$ with $\tau$ being the lifetime of the Rydberg state. The initial-state-specified fidelity is defined by $F={\rm tr}(\rho|\Psi_t\rangle\langle\Psi_t|)$ with $|\Psi_t\rangle$ being the target state. From Fig.~\ref{f6}(a) that shows the effects of different Rydberg-state lifetimes on the fidelities of the two kinds of SWAP gates, we know that the nonadiabatic holonomic SWAP gate is affected more significantly by a short Rydberg-state lifetime, while the fidelity of the dynamical one is always higher than 0.992 when $\tau>50~\mu$s, even though the gate time of the dynamical SWAP gate is prolonged by near ten multiples over that of the nonadiabatic holonomic format. For an accessible Rydberg state lifetime $\tau\sim400~\mu$s of $|r\rangle=|70S_{1/2}\rangle$~\cite{Beterov2009}, the fidelities of the two kind of SWAP gates can both reach 0.996.

\subsection{Doppler dephasing}
Due to the atomic thermal motion, processes of Rydberg excitations suffer from Doppler dephasing inevitably because of the Doppler effect~\cite{Saffman_2016}, which is an important resource of technical errors, and some works were contributed recently to suppressing Dopplor dephasing errors in implementing Rydberg-mediated quantum gates~\cite{Ryabtsev2011,Shi2019,Shi2020,JLWu2021}. When considering Doppler dephasing, detunings of the Rydberg pumping for each atom in Eq.~(\ref{e1}) are changed~\cite{Ryabtsev2011,PhysRevA.97.053803,Shi2019,Shi2020,JLWu2021}, which makes the two-atom full Hamiltonian in the interaction picture become
\begin{equation}
	\hat H'=\hat H_I+\sum_{j=1}^2\delta_j|r\rangle_j\langle r|,
\end{equation}
where $\hat H_I$ is given in Eq.~(\ref{e2}).
The added detunings $\Delta'_{1,2}$ of the Rydberg pumping lasers seen by the atoms are two random variables yielded with a Gaussian probability distribution of the mean $\bar\Delta=0$ and the standard deviation $\sigma_{\Delta} = k_{\rm eff}v_{\rm rms}$, where $k_{\rm eff}$ is the effective wave vector magnitude of lasers that atoms undergo,
and  $v_{\rm rms} = \sqrt{k_{\rm B}T_a/M}$ is the atomic root-mean-square velocity with $k_{\rm B}$, $T_a$,
and $M$ being the Boltzmann constant, atomic temperature, and atomic mass, respectively. Here we suppose for simplicity that there are two counterpropagating laser fields with wavelengths $\lambda_1\sim420$~nm and $\lambda_2 \sim1013$~nm used for Rydberg pumping~\cite{Levine2018,Bernien2017,Levine2019}, which result in an effective wave vector magnitude $k_{\rm eff}\sim8.76\times10^6~{\rm m}^{-1}$~\cite{PhysRevA.97.053803,Levine2019}. Then with these settings we numerically work out in Fig.~\ref{f6}(b) the fidelities of performing the nonadiabatic holonomic and dynamical SWAP gates, respectively, with varying the atomic temperature. The fidelity of the dynamical SWAP gate is reduced little by the Doppler dephasing, and the fidelity can be over 0.99 under $T_a\in(0,50]~\mu$K. However, the fidelity of the nonadiabatic holonomic SWAP gate is reduced relatively significantly by the Doppler dephasing. With an experimentally accessible atomic temperature $T_a\sim10~\mu{\rm K}$~\cite{Levine2019,Graham2019}, the fidelities affected by Doppler dephasing are 0.9821 and 0.9924 for the nonadiabatic holonomic and dynamical SWAP gates, respectively.

\begin{figure}[t]\centering
	\includegraphics[width=\linewidth]{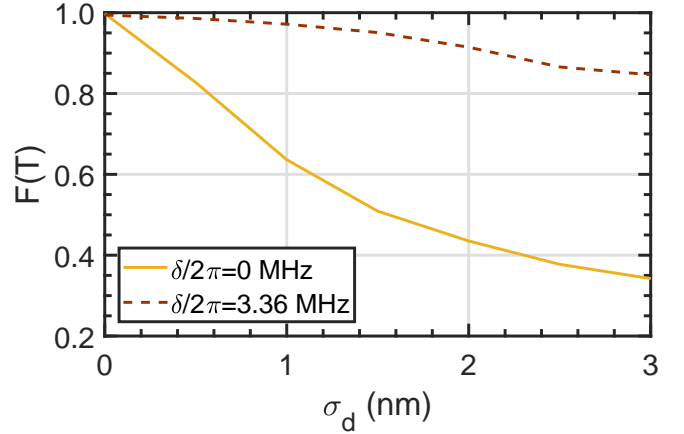}\\ 
	\caption{Fidelities of the nonadiabatic holonomic~($\delta/2\pi=0$) and dynamical~($\delta/2\pi=3.36$~MHz) SWAP gates with considering deviations of the interatomic distance. Each point denotes the average of 201 results.}\label{f7}
\end{figure}
\subsection{Deviations of interatomic distance}
It is a continuous challenge to suppress uncontrollable deviations in an ideal RRI strength due to imperfections in cooling, trapping, and manipulating atoms~\cite{Saffman_2016,Weiss2017,Browaeys2020,Henriet2020,Jo2020,Su2020}, so it is very important to assess the effect of deviations in the RRI strength $V$ on the SWAP gate schemes. Because the RRI strength $V$ changes very sharply with the interatomic distance $d$~(i.e., $V=C_6/d^6$), the performance of the SWAP gates will be very sensitive to the fluctuations of $d$. The fluctuations of $d$ can be characterized with a quasi one-dimensional Gaussian probability distribution of the mean (ideal value) $d=\sqrt[6]{C_6/V}$ and the standard deviation $\sigma_d$~\cite{Graham2019}. We simulate the effect of fluctuations in $d$ on the gate performance through plotting the fidelities of the nonadiabatic holonomic and dynamical SWAP gates, respectively, at their corresponding gate time with varying the standard deviation $\sigma_d$ of $d$. From Fig.~\ref{f7} we know that the performance of the nonadiabatic holonomic SWAP gate is significantly spoiled even when $\sigma_d< 1$~nm, which is similar to other RAB-based gate schemes~\cite{Su2016,Su2017,Su2017PRA,Su2018,Wu2020PLA,Xing2020,Wu2021PR}. By contrast, while the gate performance is still sensitive to $\sigma_d$, the robustness against the interatomic distance deviation in the dynamical SWAP gate is enhanced largely, which makes the tolerable interatomic distance fluctuations in implementing a SWAP gate closer to attainable values in the recent experiments~\cite{Graham2019,Jo2020}.

\section{Fredkin gates by combining Rydberg blockade and antiblockade}\label{sec5}
\begin{figure}[htb]\centering
	\includegraphics[width=0.88\linewidth]{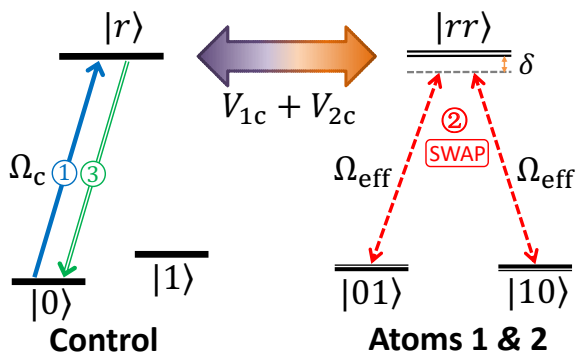}\\ 
	\caption{Schematic for implementing a Fredkin gate. A control atom is excited resonantly by a laser field from $|0\rangle_c$ to $|r\rangle_c$ with Rabi frequency $\Omega_c$. The control atom is coupled to target atoms $1$ and $2$ described in Fig.~\ref{f1}, with RRI strengths $V_{1c}$ and $V_{2c}$, respectively, which is equivalent to that the control atom is coupled to the effective $\Lambda$-type system of the target atoms that is described by Eq.~(\ref{e7}), with coupling strength $V_{1c}+V_{2c}$.}\label{f8}
\end{figure}
The Fredkin gate~\cite{Fredkin1982} is one of the most representative multi-qubit gates, swapping the quantum states of two target qubits depending on the state of a control qubit, which holds important applications in quantum error correction~\cite{Chuang1996}, quantum fingerprinting~\cite{Buhrman2001}, and quantum routers~\cite{Behera2019}. Decomposing a Fredkin gate into universal single- and two-qubit gates is pretty complicated~\cite{Feng2020}, so the direct implementation of a Fredkin gate is of great interest. In this section we propose to implement Fredkin gates by combining Rydberg blockade and antiblockade, on the basis of the nonadiabatic holonomic and dynamical SWAP gates.

Schematic for implementing Fredkin gates is shown in Fig.~\ref{f8}. The Fredkin gates based on the nonadiabatic holonomic and dynamical SWAP gates are implemented in three steps, following the idea of Jaksch et al. for implementing a two-atom Rydberg-blockade-based phase gate~\cite{Jaksch2000}.

$Step~(i)$: Apply a $\pi$ pulse to the control atom. The transformation of three-qubit computational states follows~(defining $|lmn\rangle\equiv|l\rangle_c\otimes|m\rangle_1\otimes|n\rangle_2$ with $l,m,n=0,1$)
\begin{eqnarray}\label{e12}
	&&|000\rangle\mapsto-{\rm i}|r00\rangle;~|001\rangle\mapsto-{\rm i}|r01\rangle;~|010\rangle\mapsto-{\rm i}|r10\rangle;\nonumber\\
	&&|011\rangle\mapsto-{\rm i}|r11\rangle;~|100\rangle\mapsto|100\rangle;~|101\rangle\mapsto|101\rangle;\nonumber\\
	&&|110\rangle\mapsto|110\rangle;~|111\rangle\mapsto|111\rangle.
\end{eqnarray}

$Step~(ii)$: Turn off the drive on the control atom and execute the nonadiabatic holonomic or dynamical SWAP gate operation on the target atoms 1 and 2. There are two cases needed to consider. Case 1: the state of the control atom is $|r\rangle$, which will strongly shift the energies of the Rydberg state $|r\rangle$ of the target atoms, i.e., Rydberg blockade, and hence make the SWAP operation on the target atoms work in vain under the situation $|V_{1c}+V_{2c}|\gg|\Omega_{\rm eff}|,|\delta|$. Case 2: the state of the control atom is $|1\rangle$, and the SWAP operation on the target atoms works well. Then the following transformation occurs
\begin{eqnarray}\label{e13}
	&&-{\rm i}|r00\rangle\mapsto-{\rm i}|r00\rangle;~-{\rm i}|r01\rangle\mapsto-{\rm i}|r01\rangle;\nonumber\\
	&&-{\rm i}|r10\rangle\mapsto-{\rm i}|r10\rangle;~-{\rm i}|r11\rangle\mapsto-{\rm i}|r11\rangle;\nonumber\\
	&&|100\rangle\mapsto|100\rangle;~|101\rangle\mapsto-|110\rangle;\nonumber\\
	&&|110\rangle\mapsto-|101\rangle;~|111\rangle\mapsto|111\rangle.
\end{eqnarray}

$Step~(iii)$: Apply again a $\pi$ pulse to the control atom. Therefore, the state transformation becomes finally
\begin{eqnarray}
	&&-{\rm i}|r00\rangle\mapsto-|000\rangle;~-{\rm i}|r01\rangle\mapsto-|001\rangle;\nonumber\\
	&&-{\rm i}|r10\rangle\mapsto-|010\rangle;~-{\rm i}|r11\rangle\mapsto-|011\rangle;\nonumber\\
	&&|100\rangle\mapsto|100\rangle;~|101\rangle\mapsto-|110\rangle;\nonumber\\
	&&|110\rangle\mapsto-|101\rangle;~|111\rangle\mapsto|111\rangle.
\end{eqnarray}

\begin{figure}[t]\centering
	\includegraphics[width=0.8\linewidth]{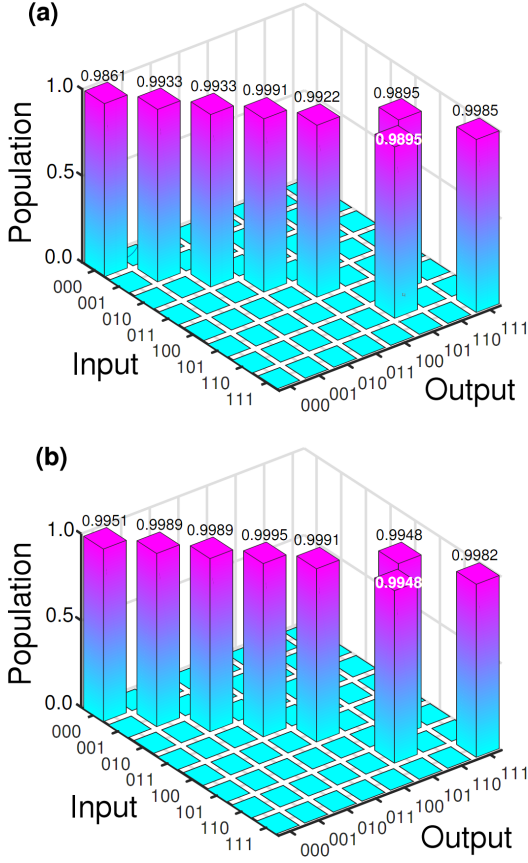}\\ 
	\caption{Truth tables of two kinds of Fredkin gates, respectively, based on (a)~the nonadiabatic holonomic SWAP gate using \{$\delta/2\pi=0$, $V_{1c}=V_{2c}=2\pi\times3$~MHz\} and (b)~the dynamical SWAP gate using \{$\delta/2\pi=3.36$~MHz, $V_{1c}=V_{2c}=2\pi\times50$~MHz\}. We set $\Omega_c/2\pi=10$~MHz, and other parameters are the same as those used for the nonadiabatic holonomic or dynamical SWAP gates.}\label{f9}
\end{figure}
The three-step procedure above combines Rydberg blockade and RAB, and ensures the implementation of a Fredkin gate based on the nonadiabatic holonomic or dynamical SWAP gate. Because the two-atom effective Rabi frequency $\Omega_{\rm eff}$ and detuning $\delta$ are relatively small, the Rydberg blockade condition $|V_{1c}+V_{2c}|\gg|\Omega_{\rm eff}|,|\delta|$ is easy to satisfy. We characterize the performance of the Fredkin gates by means of the truth tables~\cite{Xing2020,Reed2012}, which are obtained by calculating the population $|\langle\phi_{\rm out}|\mathcal{U}(T)|\phi_{\rm in}\rangle|^2$ of all the computational
states $|\phi_{\rm in}\rangle,|\phi_{\rm out}\rangle\in\{|lmn\rangle\}$~($l,m,n=0,1$), $\mathcal{U}(T)$ being the evolution operator at the gate time $t=T$. The $8\times8$ truth tables of two kinds of Fredkin gates are shown in Fig.~\ref{f9}, where the numbers on the tops of bars indicate fidelities of corresponding input states. Apparently, Fredkin gates can be achieved with high fidelities. The truth tables do not contain phase information of each computational state. Furthermore, to consider the phase information of each computational state, we evaluate the performance of Fredkin gates using the gate fidelity taken by~\cite{Souza2012,Genov2017,BJLiu2008.02176}
\begin{equation}
	F(T)=\frac18\left|{\rm tr}\left[U'(T)U_{\rm Fredkin}^\dag\right]\right|,
\end{equation}
where $U_{\rm Fredkin}=|1\rangle_c\langle1|\otimes U_{\rm SWAP}-|0\rangle_c\langle0|\otimes\hat{\mathcal{I}}_1\otimes\hat{\mathcal{I}}_2$ is the expected form of a Fredkin gate with $\hat{\mathcal{I}}$ be a $2\times2$ unit matrix on the basis $\{|0\rangle,|1\rangle\}$, while $U'(T)$ is obtained actually by solving Schr\"odinegr equation based on the three-atom full Hamiltonian. The fidelities of Fredkin gates based on the nonadiabatic holonomic and dynamical SWAP gates are 0.9944 and 0.9935, respectively.

\section{Conclusion}\label{sec6}
In summary, we have shown a dynamics regime of Rydberg atoms, unselective ground-state blockade, in which the evolution of two atoms populating in an identical ground state is blocked. We apply this regime to the implementation of a nonadiabatic holonomic SWAP gate. Furthermore, in order to circumvent common issues caused by Rydberg antiblockade, we propose to implement a dynamical SWAP gate, which does not involve the attendance of the doubly-excited Rydberg state and thus greatly strengthens the robustness of the SWAP gate. Finally, we explore to realize Fredkin gates based on the nonadiabatic holonomic and dynamical SWAP gates by means of combining the Rydberg blockade and antiblockade, and the truth tables of the Fredkin gates indicate high three-qubit gate fidelities. The present work shows an effective way to implement Rydberg antiblockade gates with the enhanced robustness, which may accelerate the experimental demonstration of quantum gates on strongly coupled atoms in the Rydberg antiblockade regime.\\

\section*{Acknowledgements}
The authors acknowledge funding from the  National Natural Science Foundation of China (NSFC) (11675046, 21973023, 11804308); Program for Innovation Research of Science in Harbin Institute of Technology (A201412); Postdoctoral Scientific Research Developmental Fund of Heilongjiang Province (LBH-Q15060); and Natural Science Foundation of Henan Province under Grant No.~202300410481.
We thank the HPC Studio at School of Physics of Harbin Institute of Technology 37 for access to computing resources through INSPUR-HPC@PHY.HIT.EDU.

\appendix
\section{Deviation of the effective Hamiltonian Eq.~(\ref{e5})}\label{Appendix}
Using the second-order perturbation theory for the full Hamiltonian Eq.~(\ref{e4}) with the condition $\Delta_{0,1}\gg|\Omega_{0,1}|$ and $\Delta_1-\Delta_0=V$, single-excitation states $|r0\rangle$, $|0r\rangle$, $|r1\rangle$, and $|1r\rangle$ can not be involved in any transitions, but have Stark shifts in energy, respectively, as
\begin{eqnarray}\label{A1}
	\Delta_{r0}&=&\frac{\langle r0|\hat{\mathcal{H}}_I|00\rangle\langle00|\hat{\mathcal{H}}_I|r0\rangle}{\Delta_0}+\frac{\langle r0|\hat{\mathcal{H}}_I|10\rangle\langle10|\hat{\mathcal{H}}_I|r0\rangle}{-\Delta_1}\nonumber\\
	&&+\frac{\langle r0|\hat{\mathcal{H}}_I|rr\rangle\langle rr|\hat{\mathcal{H}}_I|r0\rangle}{-(\Delta_0+V)}=\frac{\Omega_0^2}{4\Delta_0}-\frac{\Omega_0^2+\Omega_1^2}{4\Delta_1},\nonumber\\
	\Delta_{0r}&=&\frac{\langle 0r|\hat{\mathcal{H}}_I|00\rangle\langle00|\hat{\mathcal{H}}_I|0r\rangle}{\Delta_0}+\frac{\langle 0r|\hat{\mathcal{H}}_I|01\rangle\langle01|\hat{\mathcal{H}}_I|0r\rangle}{-\Delta_1}\nonumber\\
	&&+\frac{\langle 0r|\hat{\mathcal{H}}_I|rr\rangle\langle rr|\hat{\mathcal{H}}_I|0r\rangle}{-(\Delta_0+V)}=\frac{\Omega_0^2}{4\Delta_0}-\frac{\Omega_0^2+\Omega_1^2}{4\Delta_1},\nonumber\\
	\Delta_{r1}&=&\frac{\langle r1|\hat{\mathcal{H}}_I|01\rangle\langle01|\hat{\mathcal{H}}_I|r1\rangle}{\Delta_0}+\frac{\langle r1|\hat{\mathcal{H}}_I|11\rangle\langle11|\hat{\mathcal{H}}_I|r1\rangle}{-\Delta_1}\nonumber\\
	&&+\frac{\langle r1|\hat{\mathcal{H}}_I|rr\rangle\langle rr|\hat{\mathcal{H}}_I|r1\rangle}{\Delta_1-V}=\frac{\Omega_0^2+\Omega_1^2}{4\Delta_0}-\frac{\Omega_0^2}{4\Delta_1},\nonumber\\
	\Delta_{1r}&=&\frac{\langle 1r|\hat{\mathcal{H}}_I|10\rangle\langle10|\hat{\mathcal{H}}_I|1r\rangle}{\Delta_0}+\frac{\langle 1r|\hat{\mathcal{H}}_I|11\rangle\langle11|\hat{\mathcal{H}}_I|1r\rangle}{-\Delta_1}\nonumber\\
	\cr&&+\frac{\langle 1r|\hat{\mathcal{H}}_I|rr\rangle\langle rr|\hat{\mathcal{H}}_I|1r\rangle}{\Delta_1-V}=\frac{\Omega_0^2+\Omega_1^2}{4\Delta_0}-\frac{\Omega_0^2}{4\Delta_1}.\nonumber\\
\end{eqnarray}
These Stark shifts can be dropped safely because the single-excitation states are decoupled to the four two-atom ground states that are set to be the initial state of the system.  

From Eq.~(\ref{e4}) and Fig.~\ref{f2} we can clearly find that the two-photon transition between $|00\rangle$~($|11\rangle$) and $|rr\rangle$ mediated by $|0r\rangle$ or $|r0\rangle$~($|1r\rangle$ or $|r1\rangle$) is far off-resonant with a large blue detuning $2\Delta_0+V$~(red detuning $2\Delta_1-V$), which is the very reason why the evolution of $|00\rangle$ or $|11\rangle$ is blocked. By contrast,  the two-photon transition between $|01\rangle$~($|10\rangle$) and $|rr\rangle$ mediated by $|0r\rangle$ or $|r1\rangle$~($|r0\rangle$ or $|1r\rangle$) is exactly resonant under the condition $\Delta_1-\Delta_0=V$. Effective Rabi frequencies $\Omega_{\rm eff01}$ and $\Omega_{\rm eff01}$ of the two-photon transitions $|01\rangle\leftrightarrow|rr\rangle$ and $|10\rangle\leftrightarrow|rr\rangle$ can be worked out, respectively, by
\begin{eqnarray}
	\frac{\Omega_{\rm eff01}}2&=&\frac{\langle 01|\hat{\mathcal{H}}_I|0r\rangle\langle0r|\hat{\mathcal{H}}_I|01\rangle}{\Delta_1}+\frac{\langle 01|\hat{\mathcal{H}}_I|r1\rangle\langle r1|\hat{\mathcal{H}}_I|01\rangle}{-\Delta_0}\nonumber\\
	&=&\frac{\Omega_0\Omega_1}{4\Delta_1}-\frac{\Omega_0\Omega_1}{4\Delta_0},\nonumber\\
	\frac{\Omega_{\rm eff10}}2&=&\frac{\langle 10|\hat{\mathcal{H}}_I|r0\rangle\langle r0|\hat{\mathcal{H}}_I|10\rangle}{\Delta_1}+\frac{\langle 10|\hat{\mathcal{H}}_I|1r\rangle\langle 1r|\hat{\mathcal{H}}_I|10\rangle}{-\Delta_0}\nonumber\\
	&=&\frac{\Omega_0\Omega_1}{4\Delta_1}-\frac{\Omega_0\Omega_1}{4\Delta_0}.
\end{eqnarray}

In addition to the effective coupling of $|rr\rangle$ to $|01\rangle$ and $|10\rangle$, there are Stark shifts in energy of $|00\rangle$, $|01\rangle$, $|10\rangle$, $|11\rangle$, and $|rr\rangle$ calculated out, respectively, as
\begin{eqnarray}
	\Delta_{00}&=&\frac{\langle 00|\hat{\mathcal{H}}_I|r0\rangle\langle r0|\hat{\mathcal{H}}_I|00\rangle}{-\Delta_0}+\frac{\langle 00|\hat{\mathcal{H}}_I|0r\rangle\langle 0r|\hat{\mathcal{H}}_I|00\rangle}{-\Delta_0}\nonumber\\
	&=&\frac{-\Omega_0^2}{2\Delta_0},\nonumber\\
	\Delta_{01}&=&\frac{\langle 01|\hat{\mathcal{H}}_I|r1\rangle\langle r1|\hat{\mathcal{H}}_I|01\rangle}{-\Delta_0}+\frac{\langle 01|\hat{\mathcal{H}}_I|0r\rangle\langle 0r|\hat{\mathcal{H}}_I|01\rangle}{\Delta_1}\nonumber\\
	&=&\frac{\Omega_1^2}{4\Delta_1}-\frac{\Omega_0^2}{4\Delta_0},\nonumber\\
	\Delta_{10}&=&\frac{\langle 10|\hat{\mathcal{H}}_I|1r\rangle\langle 1r|\hat{\mathcal{H}}_I|10\rangle}{-\Delta_0}+\frac{\langle 10|\hat{\mathcal{H}}_I|r0\rangle\langle r0|\hat{\mathcal{H}}_I|10\rangle}{\Delta_1}\nonumber\\
	&=&\frac{\Omega_1^2}{4\Delta_1}-\frac{\Omega_0^2}{4\Delta_0},\nonumber\\
	\Delta_{11}&=&\frac{\langle 11|\hat{\mathcal{H}}_I|r1\rangle\langle r1|\hat{\mathcal{H}}_I|11\rangle}{\Delta_1}+\frac{\langle 11|\hat{\mathcal{H}}_I|1r\rangle\langle 1r|\hat{\mathcal{H}}_I|11\rangle}{\Delta_1}\nonumber\\
	&=&\frac{\Omega_1^2}{2\Delta_1},\nonumber\\
	\cr\Delta_{rr}&=&\frac{\langle rr|\hat{\mathcal{H}}_I|r0\rangle\langle r0|\hat{\mathcal{H}}_I|rr\rangle}{V+\Delta_0}+\frac{\langle rr|\hat{\mathcal{H}}_I|0r\rangle\langle 0r|\hat{\mathcal{H}}_I|rr\rangle}{V+\Delta_0}\nonumber\\
	&&+\frac{\langle rr|\hat{\mathcal{H}}_I|r1\rangle\langle r1|\hat{\mathcal{H}}_I|rr\rangle}{-(\Delta_1-V)}+\frac{\langle rr|\hat{\mathcal{H}}_I|1r\rangle\langle 1r|\hat{\mathcal{H}}_I|rr\rangle}{-(\Delta_1-V)}\nonumber\\
	&=&\frac{\Omega_0^2}{2\Delta_1}-\frac{\Omega_1^2}{2\Delta_0}.
\end{eqnarray}
According to the analysis above, an effective form of Hamiltonian Eq.~(\ref{e4}) is obtained exactly as Eq.~(\ref{e5}).

\section{Deviation of the effective Hamiltonian Eq.~(\ref{e7})}\label{A2}
There are several methods eliminating Stark shifts, such as phase	corrections~\cite{Veps_l_inen_2018,Vepsaaineneaau5999}, auxiliary detuned transitions~\cite{Su2016,YJZhao2017}, and optimal search of detunings~\cite{Wang_2020,Han2020}. According to the second-order perturbation theory, the auxiliary atom-field interaction for the two atoms induces the Stark shifts on the ground states as
$\sum_{j=1}^2\sum_{k=0}^1(-1)^k{\Omega'_k}^2/4\Delta'_k|k\rangle_j\langle k|$ that can be rewritten on the two-atom basis
\begin{eqnarray}
	&&\frac{{\Omega'_0}^2}{4\Delta'_0}(2|00\rangle\langle 00| +|0r\rangle\langle 0r|+|r0\rangle\langle r0|)\nonumber\\
	&&+\left(\frac{{\Omega'_0}^2}{4\Delta'_0}-\frac{{\Omega'_1}^2}{4\Delta'_1}\right)(|01\rangle\langle 01|+|10\rangle\langle 10|)
	\nonumber\\
	&&-\frac{{\Omega'_1}^2}{4\Delta'_1}(2|11\rangle\langle 11| +|1r\rangle\langle 1r|+|r1\rangle\langle r1|).
\end{eqnarray}
Apparently, these Stark shifts can exactly counteract the ground-state Stark shifts in Eq.~(\ref{e5}) by setting the relation ${{\Omega'_k}^2}/{\Delta'_k}={{\Omega_k}^2}/{\Delta_k}$~($k=1,2$), and those for the single-excitation states can be dropped because they are decoupled to the four ground states.

For dealing with the Stark shift of $|rr\rangle$, we transform Hamiltonian Eq.~(\ref{e2}) into the frame defined by a rotation operator $U'_0=\exp[{\rm i}t(\Delta_1-\Delta_0)|rr\rangle\langle rr|]$ instead of $U_0=\exp({\rm i}tV|rr\rangle\langle rr|)$ used for Eq.~(\ref{e4}), so we obtain
\begin{eqnarray}
	&&\hat{\mathcal{H}}'_I=\nonumber\\
	&&\Big[\frac{\Omega_0}2e^{-{\rm i}\Delta_0t}(|00\rangle\langle r0|+|01\rangle\langle r1|+|00\rangle\langle 0r|+|10\rangle\langle 1r|)\nonumber\\
	&&+\frac{\Omega_1}2e^{{\rm i}\Delta_1t}(|10\rangle\langle r0|+|11\rangle\langle r1|+|01\rangle\langle 0r|+|11\rangle\langle 1r|)\nonumber\\
	&&+\frac{\Omega_0}2e^{-{\rm i}\Delta_1t}(|0r\rangle\langle rr|+|r0\rangle\langle rr|)+\frac{\Omega_1}2e^{{\rm i}\Delta_0t}(|1r\rangle\langle rr|\nonumber\\
	&&+|r1\rangle\langle rr|)+{\rm H.c.}\Big]+V_0|rr\rangle\langle rr|,
\end{eqnarray}
where we have defined a small quantity $V_0=V-(\Delta_1-\Delta_0)$. Then an effective Hamiltonian is of the form $\hat H_{\rm e}=\hat{\mathcal{H}}'_I+V_0|rr\rangle\langle rr|$, where $\hat{\mathcal{H}}'_I$ is given in Eq.~(\ref{e5}). After eliminating the ground-state Stark shifts by the auxiliary atom-field interaction, a final effective Hamiltonian can be determined
\begin{eqnarray}
	\hat H_{\rm eff}&=&\Big[\frac{\Omega_{\rm eff}}2(|01\rangle\langle rr|+|10\rangle\langle rr|)+{\rm H.c.}\Big]\nonumber\\
	&&+(V_0+\Delta_{rr})|rr\rangle\langle rr|.
\end{eqnarray}

\end{document}